\documentclass{kluwer}
\usepackage{graphics}
\newdisplay{guess}{Conjecture}

\begin{document}
\begin{article}
\begin{opening}

 \title{Tensor and spin representations of SO(4) and discrete quantum gravity }

 \author{Miguel \surname{Lorente}}
 \runningauthor{Miguel Lorente and Peter Kramer}
 \runningtitle{Tensor and spin representations of SO(4) 
 and discrete quantum gravity}
 \institute{Departamento de F\'{\i}sica, Universidad de Oviedo, 
 33007 Oviedo, Spain}
 \author{Peter Kramer}
 \institute{Institut f\" ur theoretische Physik Universit\" at T\" ubingen, 72076 T\" ubingen, Germany}
\date{October 27, 2003}

\begin{abstract}
Starting from the defining transformations of complex matrices for the $SO(4)$ group, we construct the fundamental
representation and the tensor and spinor representations of the group $SO(4)$. Given the commutation relations for the 
corresponding algebra, the unitary representations of the group in terms of the generalized Euler angles are
constructed. These mathematical results help us to a more complete description of the Barrett-Crane model in Quantum
Gravity. In particular a complete realization of the weight function for the partition function is given and a new
geometrical interpretation of the asymptotic limit for the Regge action is presented. 
\end{abstract}


\end{opening}    

\section{Discrete models in quantum gravity}

The use of discrete models in Physics has become very popular, mainly for two reasons. It helps to find the solutions of
some differential equations by numerical methods, which would not be possible to solve by analytic methods. Besides that,
the introduction of a lattice is equivalent to the introduction of a cut-off in the momentum variable for the field in
order to achieve the finite limit of the solution. In the case of relativistic field equations -like the Dirac,
Klein-Gordon, and the electromagnetic interactions- we have worked 
out some particular cases [1].

There is an other motivation for the discrete models and it is based in some philosophical presuppositions that the
space-time structure is discrete. This is more attractive in the case of general relativity and quantum gravity because it
makes more transparent the connection between the discrete properties of the intrinsic curvature and the background
independent gravitational field.

This last approach was started rigorously by Regge in the early sixties [2]. He introduces some triangulation in a
Riemannian manifold, out of which he constructs local curvature, 
coordinate independent, on the polyhedra. With the help of
the total curvature on the vertices of the discrete manifold he 
constructs a finite action which, in the continuous limit,
becomes the standard Hilbert-Einstein action of general relativity.

Regge himself applied his method (``Regge calculus'') to quantum gravity in three dimensions [3]. In this work he assigns
some representation of the $SU(2)$ group to the edges of the triangles. To be more precise, to every tetrahedron appearing in
the discrete triangulation of the manifold he associates a 6j-symbol in such a way that the spin eigenvalues of the
corresponding representation satisfy sum rules described by the edges and vertices of the tetrahedra. Since the value of
the 6j-symbol has a continuous limit when some edges of the tetrahedra become very large, he could calculate the sum of
this limit for all the 6j-symbols attached to the tetrahedra, and in this way he could compare it with the continuous
Hilbert-Einstein action corresponding to an Euclidean non planar manifold.

A different approach to the discretization of space and time was taken by Penrose [4]. Given some graph representing the
interaction of elementary units satisfying the rules of angular momentum 
without an underlying space, he constructs out of this
network (``spin network'') the properties of total angular momentum as a derived concept. Later this model was applied to
quantum gravity in the sense of Ponzano and Regge. In general, a spin network is a triple $(\gamma, \rho, i)$ where 
$\gamma$ is a graph with a finite set of edges $e$, a finite set of vertices $v$, $\rho_e$ is the representation of a
group G attached to an edge, and $i_v$ is an intertwiner attached to each vertex. If we take the product of the amplitudes
corresponding to all the edges and vertices (given in terms of the representations and intertwiners) we obtain the
particular diagram of some quantum state.

Although the physical consequences of Penrose's ideas were soon considered  to be equivalent to the Ponzano-Regge approach to
quantum gravity [5], the last method was taken as guiding rule in the calculation of partition functions. We can mention a few
results. Turaev and Viro [6] calculated the state sum for a 3d-triangulated manifold with tetrahedra described by 6j-symbols
using the $SU(2)_q$ group. This model was enlarged to 4-dimensional triangulations and was proved by Turaev, Oguri, Crane and
Yetter [7] to be independent of the triangulation (the ``TOCY model'').

A different approach was introduced by Boulatov [8] that led to the same partition function as the TOCY model, but with
the advantage that the terms corresponding to the kinematics and the interaction could be distinguished. For this purpose
he introduced some fields defined over the elements of the groups $SO(3)$, invariant under the action of the group, and
attached to the edges of the tetrahedra. The kine\-ma\-ti\-cal term corresponds to the self interacting field over each edge and
the interaction term corresponds to the fields defined in different edges and coupled among themselves. This method (the
Boulatov matrix model) was very soon enlarged to 4-dimensional triangulations by Ooguri [9]. In both models the fields over
the matrix elements of the group are expanded in terms of the representations of the group and then integrated out, with
the result of a partition function extended to the amplitudes over all tetrahedra, all edges and vertices of the
triangulation.

A more abstract approach was taken by Barrett and Crane, generalizing Penrose's spin networks to 4 dimensions. The novelty
of this model consists in the association of representation of the $SO(4)$ group to the faces of the tetrahedra. We will
come back to this model in section 5.

Because we are interested in the physical and mathematical properties of the Barrett-Crane model, we mention briefly some
recent work about this model combined with the matrix model approach of Boulatov and Ooguri [10]. In this work the 2d--
quantum space-time emerges as a  Feynman graph, in the manner of the 4d-- matrix models. In this way a spin foam model is
connected to the Feynman diagram of quantum gravity.

In this paper we have tried to implement all the mathematical consequences of Barrett-Crane model using the group
theoretical properties of $SO(4)$ applied to the 4d-triangulation of manifolds in terms of 4-simplices. It turns out that when
we take into account the description of spin representations of $SO(4)$ the weight function given by Barrett and Williams is
incomplete; besides the values for the areas in the Regge action can be calculated in our paper directly from geometrical
considerations.

\section{The groups ${\bf SO(4,R)}$ and ${\bf SU(2) \times SU(2)}$}

The rotation group in 4 dimensions is the group of linear transformations that leaves the quadratic form
$x_1^2+x_2^2+x_3^2+x_4^2$ invariant. The
well known fact that this group is locally isomorphic to $SU(2) \times SU(2)$ enables one to decompose the group action in
the following way:

Take a complex matrix (not necessarily unimodular)
\begin{equation}
    w=\left( {\matrix{y&z\cr
{-\bar z}&{\bar y}\cr}} \right) \quad ,\quad y=x_1+ix_2,-\bar z=x_3+ix_4,
\end{equation}
where $w$ satisfies $w\,w^+=\det (w)$.

We define the {\it complete} group action 
\begin{equation}
    w\to w'=u_1wu_2,
\end{equation}
where $u_1,u_2\in SU(2)$ correspond to the left, right action, respectively,
\begin{eqnarray*}
   u_1&=&\left( {\matrix{\alpha &\beta \cr {-\bar \beta }&{\bar \alpha }\cr}} \right) \in SU(2)^L\;\;,\;\;\alpha \bar \alpha +\beta \bar
\beta =1,\\
u_2&=&\left( {\matrix{\gamma &\delta \cr
{-\bar \delta }&{\bar \gamma }\cr
}} \right) \in SU(2)^R\;\;,\;\;\gamma \bar \gamma +\delta \bar \delta =1.
\end{eqnarray*}
The complete group action satisfies:
\begin{equation}
    w'\,w'^+=\det (w')=w\,w^+=\det (w),
\end{equation}
or $\;\; {x'_1}^2+{x'_2}^2+{x'_3}^2+{x'_4}^2={x_1}^2+{x_2}^2+{x_3}^2+{x_4}^2 \;$, which corresponds to the defining relation for $SO(4,R)$.

In order to make connection with $R^4$, we take only the {\em left} action $w'=u_1w$ and express the matrix elements of
$w$ as a 4-vector
\begin{equation}
   \left( {\matrix{{y'}\cr
{-\bar z'}\cr
{z'}\cr
{\bar y'}\cr
}} \right)=\left( {\matrix{\alpha &\beta &0&0\cr
{-\bar \beta }&{\bar \alpha }&0&0\cr
0&0&\alpha &\beta \cr
0&0&{-\bar \beta }&{\bar \alpha }\cr
}} \right)\left( {\matrix{y\cr
{-\bar z}\cr
z\cr
{\bar y}\cr
}} \right).
\end{equation}

Substituting $y=x_1+ix_2\;, -\bar z=x_3+ix_4$, and $\alpha =\alpha _1+i\alpha _2\;,\;\beta =\beta _1+i\beta _2$, we get
\begin{equation}
    \left( {\matrix{{x'_1}\cr
{x'_2}\cr
{x'_3}\cr
{x'_4}\cr
}} \right)=\left( {\matrix{{\alpha _1}&{-\alpha _2}&{\beta _1}&{-\beta _2}\cr
{\alpha _2}&{\alpha _1}&{\beta _2}&{\beta _1}\cr
{-\beta _1}&{-\beta _2}&{\alpha _1}&{\alpha _2}\cr
{\beta _2}&{-\beta _1}&{-\alpha _2}&{\alpha _1}\cr
}} \right)\left( {\matrix{{x_1}\cr
{x_2}\cr
{x_3}\cr
{x_4}\cr
}} \right).
\end{equation}

Obviously, the transformation matrix is orthogonal. Similarly for the right action $w'=wu_2^+$
 we get
\begin{equation}
    \left( {\matrix{{y'}\cr
{-\bar z'}\cr
{z'}\cr
{\bar y'}\cr
}} \right)=\left( {\matrix{{\bar \gamma }&0&{\bar \delta }&0\cr
0&{\bar \gamma }&0&{\bar \delta }\cr
{-\delta }&0&\gamma &0\cr
0&{-\delta }&0&\gamma \cr
}} \right)\left( {\matrix{y\cr
{-\bar z}\cr
z\cr
{\bar y}\cr
}} \right),
\end{equation}
and after substituting $\gamma =\gamma _1+i\gamma _2\;,\;\delta =\delta _1+i\delta _2$, we get
\begin{equation}
   \left( {\matrix{{x'_1}\cr
{x'_2}\cr
{x'_3}\cr
{x'_4}\cr
}} \right)=\left( {\matrix{{\gamma _1}&{\gamma _2}&{-\delta _1}&{\delta _2}\cr
{-\gamma _2}&{\gamma _1}&{\delta _2}&{\delta _1}\cr
{\delta _1}&{-\delta _2}&{\gamma _1}&{\gamma _2}\cr
{-\delta _2}&{-\delta _1}&{-\gamma _2}&{\gamma _1}\cr
}} \right)\left( {\matrix{{x_1}\cr
{x_2}\cr
{x_3}\cr
{x_4}\cr
}} \right),
\end{equation}
where the transformation matrix is orthogonal.

If we take the complete action
\[
   \left( {\matrix{{y'}&{z'}\cr
{-\bar z'}&{\bar y'}\cr
}} \right)=\left( {\matrix{\alpha &\beta \cr
{-\bar \beta }&{\bar \alpha }\cr
}} \right)\left( {\matrix{y&z\cr
{-\bar z}&{\bar y}\cr
}} \right)\left( {\matrix{{\bar \gamma }&{-\delta }\cr
{\bar \delta }&\gamma \cr
}} \right),
\]
we get
\begin{eqnarray}
  \nonumber \left( {\matrix{{y'}\cr
{-\bar z'}\cr
{z'}\cr
{\bar y'}\cr
}} \right)&=&\left( {\matrix{{\alpha \bar \gamma }&{\beta \bar \gamma }&{\alpha \bar \delta }&{\beta \bar \delta }\cr
{-\bar \beta \bar \gamma }&{\bar \alpha \bar \gamma }&{-\bar \beta \bar \delta }&{\bar \alpha \bar \delta }\cr
{-\alpha \delta }&{-\beta \delta }&{\alpha \gamma }&{\beta \gamma }\cr
{\bar \beta \delta }&{-\bar \alpha \delta }&{-\bar \beta \gamma }&{\bar \alpha \gamma }\cr
}} \right)\left( {\matrix{y\cr
{-\bar z}\cr
z\cr
{\bar y}\cr
}} \right)=\\
&=&\left( {\matrix{\alpha &\beta &0&0\cr
{-\bar \beta }&{\bar \alpha }&0&0\cr
0&0&\alpha &\beta \cr
0&0&{-\bar \beta }&{\bar \alpha }\cr
}} \right)\left( {\matrix{{\bar \gamma }&0&{\bar \delta }&0\cr
0&{\bar \gamma }&0&{\bar \delta }\cr
{-\delta }&0&\gamma &0\cr
0&{-\delta }&0&\gamma \cr
}} \right)\left( {\matrix{y\cr
{-\bar z}\cr
z\cr
{\bar y}\cr
}} \right),
\end{eqnarray}
and taking $y=x_1+ix_2\;,\;-\bar z=x_3+ix_4$ we get the general transformation matrix for the 4-dimensional vector in $R^4$
under the group $SO(4,R)$ as
\begin{equation}
\left( {\matrix{{x'_1}\cr
{x'_2}\cr
{x'_3}\cr
{x'_4}\cr
}} \right)=\left( {\matrix{{\alpha _1}&{-\alpha _2}&{\beta _1}&{-\beta _2}\cr
{\alpha _2}&{\alpha _1}&{\beta _2}&{\beta _1}\cr
{-\beta _1}&{-\beta _2}&{\alpha _1}&{\alpha _2}\cr
{\beta _2}&{-\beta _1}&{-\alpha _2}&{\alpha _1}\cr
}} \right)\left( {\matrix{{\gamma _1}&{\gamma _2}&{-\delta _1}&{\delta _2}\cr
{-\gamma _2}&{\gamma _1}&{\delta _2}&{\delta _1}\cr
{\delta _1}&{-\delta _2}&{\gamma _1}&{\gamma _2}\cr
{-\delta _2}&{-\delta _1}&{-\gamma _2}&{\gamma _1}\cr
}} \right)\left( {\matrix{{x_1}\cr
{x_2}\cr
{x_3}\cr
{x_4}\cr
}} \right).
\end{equation}

Notice that the eight parameters $\alpha _1,\alpha _2,\beta _1,\beta _2,\gamma _1,\gamma _2,\delta _1,\delta _2$ with the
constraints $\alpha _1^2+\alpha _2^2+\beta _1^2+\beta _2^2=1\;\,,\;\,\gamma _1^2+\gamma _2^2+\delta _1^2+\delta _2^2=1$,
can be considered the Cayley parameters for the $SO(4)$ group [11].

\section{Tensor and spinor representations of SO(4,R)}

Given the fundamental 4-dimensional representation of $SO(4,R)$ in terms of the parameters $\alpha ,\beta ,\gamma ,\delta
$, as given in (9),
\begin{equation}
x'_\mu =g_{\mu \nu }x_\nu, 
\end{equation}
the tensor representations are defined in the usual way
\begin{eqnarray}
T_{k'_1k'_2\ldots k'_n}=g_{k'_1k_1}&\ldots& g_{k'_nk_n}T_{k_1k_2\ldots k_n},\\
\nonumber &&\left( {k'_i,k_i=1,2,3,4} \right).
\end{eqnarray}

For the sake of simplicity we take the second rank tensors. We can decompose them into totally symmetric and antisymmetric
tensors, namely,
\vspace{-0,2cm}
\begin{eqnarray*}S_{ij}&\equiv& x_iy_j+x_jy_i \qquad \mbox{(totally symmetric)},\\
A_{ij}&\equiv& x_iy_j-x_jy_i \qquad \mbox{(antisymmetric)}.
\end{eqnarray*}
If we substract the trace from $S_{ij}$ we get a tensor that transforms under an irreducible representation. For the
antisymmetrie tensor the situation is more delicate. In general we have 
\begin{equation}
A'_{ij}\equiv x'_iy'_j-x'_jy'_i=\left( {g_{i\ell}g_{jm}-g_{j\ell}g_{im}} \right)A_{\ell m}.
\end{equation}

This representation of dimension 6 is still reducible. For simplicity take the left action of the group given in (5). The
linear combination of the antisymmetric tensor components are transformed among themselves in the following way:
\begin{equation}
\left( {\matrix{{A'_{12}+A'_{34}}\cr
{A'_{31}+A'_{24}}\cr
{A'_{23}+A'_{14}}\cr
}} \right)=\left( {\matrix{{A_{12}+A_{34}}\cr
{A_{31}+A_{24}}\cr
{A_{23}+A_{14}}\cr
}} \right),
\end{equation}
\vspace{-5mm}
\begin{eqnarray}
\nonumber &&\left( {\matrix{{A'_{12}-A'_{34}}\cr
{A'_{31}-A'_{24}}\cr
{A'_{23}-A'_{14}}\cr
}} \right)=\\
\nonumber &=&\left( {\matrix{{\alpha _1^2+\alpha _2^2-\beta _1^2-\beta _2^2}&{-2\left( {\alpha _1\beta _2-\alpha _2\beta
_1} \right)}&{-2\left( {\alpha _1\beta _1+\alpha _2\beta _2} \right)}\cr {2\left( {\alpha _1\beta _2+\alpha _2\beta _1}
\right)}&{\alpha _1^2-\alpha _2^2+\beta _1^2-\beta _2^2}&{2\left( {\alpha _1\alpha _2-\beta _1\beta _2} \right)}\cr
{2\left( {\alpha _1\beta _1-\alpha _2\beta _2} \right)}&{-2\left( {\alpha _1\alpha _2+\beta _1\beta _2} \right)}&{\alpha
_1^2-\alpha _2^2-\beta _1^2+\beta _2^2}\cr }} \right) \times \\
&\times &\left( {\matrix{{A_{12}-A_{34}}\cr {A_{31}-A_{24}}\cr
{A_{23}-A_{14}}\cr
}} \right).
\end{eqnarray}
In the case of the right action given by (7) the 6-dimensional representation for the antisymmetrie second rank tensor
decomposes into two irreducible 3-dimensional representation of $SO(4,R)$. 
For this purpose one takes the linear
combination of the components of the antisymmetric tensor as before:
\begin{equation}
\left( {\matrix{{A'_{23}-A'_{14}}\cr
{A'_{31}-A'_{24}}\cr
{A'_{12}-A'_{34}}\cr
}} \right)=\left( {\matrix{{A_{23}-A_{14}}\cr
{A_{31}-A_{24}}\cr
{A_{12}-A_{34}}\cr
}} \right),
\end{equation}
\vspace{-5mm}
\begin{eqnarray}
\nonumber &&\left( {\matrix{{A'_{23}+A'_{14}}\cr
{A'_{31}+A'_{24}}\cr
{A'_{12}+A'_{34}}\cr
}} \right)=\\
\nonumber &=&\left( {\matrix{{\gamma _1^2-\gamma _2^2-\delta _1^2+\delta _2^2}&{2\left( {\gamma _1\gamma _2+\delta _1\delta
_2} \right)}&{-2\left( {\gamma _1\delta _1-\gamma _2\delta _2} \right)}\cr {-2\left( {\gamma _1\gamma _2-\delta _1\delta
_2} \right)}&{\gamma _1^2-\gamma _2^2+\delta _1^2-\delta _2^2}&{2\left( {\gamma _1\delta _2+\gamma _2\delta _1}
\right)}\cr {2\left( {\gamma _1\delta _1+\gamma _2\delta _2} \right)}&{-2\left( {\gamma _1\delta _2-\gamma _2\delta _1} \right)}&{\gamma _1^2+\gamma _2^2-\delta _1^2-\delta _2^2}\cr }}
\right) \times \\
&\times &\left( {\matrix{{A_{23}+A_{14}}\cr {A_{31}+A_{24}}\cr
{A_{12}+A_{34}}\cr
}} \right).
\end{eqnarray}
Therefore the 6-dimensional representation for the antisymmetric tensor decomposes into two irreducible 3-dimensional
irreducible representation of the $SO(4,R)$ group.

For the spinor representation of $SU(2)^L$ we take 
\begin{equation}
\left( {\matrix{{a'_1}\cr
{a'_2}\cr
}} \right)=\left( {\matrix{\alpha &\beta \cr
{-\bar \beta }&{\bar \alpha }\cr
}} \right)\left( {\matrix{{a_1}\cr
{a_2}\cr
}} \right)\;,\quad a_1,a_2\in {\not\subset}
\end{equation}

Let $a^{i_1i_2\ldots i_k}\quad ,\quad \left( {i_1,i_2,\ldots i_k=1,2} \right)$ 
be a set of complex numbers of dimension $2^k$ which
transform under the $SU(2)^L$  group as follows:
\begin{equation}
a^{i'_1\ldots i'_k}=u_{i'_1i_1}\ldots u_{i'_ki_k}a^{i_1\ldots i_k},
\end{equation}
where $u_{i'_1i_1},u_{i'_2i_2}\ldots $ are the components of $u\in SO(2)^L$. 
If $a^{i_1\ldots i_k}$ is totally symmetric in the indices
$i_1\ldots i_k$ the representation of dimension
$(k+1)$ is irreducible. In an analogous way we can define an irreducible representation of $SU(2)^R$ with respect to the totally symmetric
multispinor of dimension $(\ell+1)$.

For the general group $SU(4,R)\sim SU(2)^L\otimes SU(2)^R$ we can take a set of totaly symmetrie multispinors that transform under the
$SO(4)$ group as
\begin{equation}
a^{i'_1\ldots i'_k\,j'_1\ldots j'_\ell }=u_{i'_1i_1}\ldots u_{i'_ki_k}\bar v _{j'_1j_1}
\ldots \bar v _{j'_\ell i_\ell} a^{i_1\ldots i_kj_1\ldots j_\ell }
\end{equation}
where $u_{i'_1i_1}\ldots $ are the components of a general element of $SU(2)^L$ and 
$\bar v _{j'_\ell i_\ell}$ are the components
of a general element of $SU(2)^R$. They define an irreducible representation of $SO(4,R)$ of dimension $(k+1)(\ell +1)$ and with labels
(see next section)
\begin{equation}
\ell _0={{k-\ell} \over 2}\;\;,\quad \ell _1={{k+\ell} \over 2}+1.
\end{equation}

\section{Representations of the algebra ${\bf so(4,R)}$}

Let $J_1,J_2,J_3$ be the generators corresponding to the rotations in the planes $(x_2,x_3),(x_3,x_1)$, and $(x_1,x_2)$ respectively, and
$K_1,K_2,K_3$ the generators corresponding to the 
rotations (boost) in the planes $(x_1,x_4)$, $(x_2,x_4)$ and $(x_3,x_4)$ respectively. They
satisfy the following conmutation relations:
\begin{eqnarray}
\nonumber &&\left[ {J_p,J_q} \right]=i\varepsilon _{pqr}J_r\;\quad ,\quad \;p,q,r=1,2,3,\\
\nonumber &&\left[ {J_p,K_q} \right]=i\varepsilon _{pqr}K_r,\\
&&\left[ {K_p,K_q} \right]=i\varepsilon _{pqr}J_r.
\end{eqnarray}
\medskip 
If one defines 
$\bar A={1 \over 2}\left( {\bar J+\bar K} \right)\quad ,
\quad \bar B={1 \over 2}\left( {\bar J-\bar K} \right),$
		
\noindent with \qquad \qquad   $\bar
J=\left( {J_1,J_2,J_3} \right)\quad ,\quad \bar K=\left( {K_1,K_2,K_3} \right)$, then 
\begin{eqnarray}
\nonumber && \left[ {A_p,A_q} \right]=i\varepsilon _{pqr}A_r\;
\quad ,\quad \;p,q,r=1,2,3,\\
\nonumber && \left[ {B_p,B_q} \right]=i\varepsilon _{pqr}B_r,\\
&& \left[ {A_p,B_q} \right]=0,
\end{eqnarray}
that is to say, the algebra so(4) decomposes into two simple algebras su(2) $\times$ su(2)

Let $\phi _{m_1m_2}$ be a basis where $\bar A^2,A_3$ and $\bar B^2,B_3$ are 
diagonal. Then a unitary irreducible representation for the
sets $\left\{ {A_\pm \equiv A_1 \pm iA_2,A_3} \right\}$ and $\left\{ {B_\pm \equiv B_1 \pm iB_2,B_3} \right\}$ is given by 
\begin{eqnarray}
\nonumber A_\pm \phi _{m_1m_2}&=&\sqrt {\left( {j_1\mp m_1} \right)\left( {j_1\pm m_1+1} \right)}\phi _{m_1\pm 1,m_2},\\
A_3\phi _{m_1m_2}&=&m_1\phi _{m_1m_2}\quad ,\quad -j_1\le m_1\le j_1,
\end{eqnarray}
\begin{eqnarray*}
B_\pm \phi _{m_1m_2}&=&\sqrt {\left( {j_2\mp m_2} \right)
\left( {j_2\pm m_2+1} \right)}\phi _{m_1m_2\pm 1},\\
B_3\phi _{m_1m_2}&=&m_2\phi _{m_1m_2}\quad ,\quad -j_2\le m_2\le j_2.
\end{eqnarray*}
We change now to a new basis
\begin{equation}
\psi _{jm}=\sum\limits_{m_1+m_2=m} \left\langle {{j_1m_1j_2m_2}} \mathrel{\left | {\vphantom {{j_1m_1j_2m_2} {jm}}} \right.
\kern-\nulldelimiterspace} {{jm}} \right\rangle \phi _{m_1m_2}
\end{equation}
that corresponds to the Gelfand-Zetlin basis for so(4), 
\[
\psi _{jm}=\left| {\matrix{{j_1+j_2}&,&{j_1-j_2}\cr
{}&j&{}\cr
{}&m&{}\cr
}} \right\rangle.
\]

In this basis the representation for the generators $\bar J,\bar K$ of so(4) 
are given by [12]
\begin{eqnarray}
\nonumber J_\pm \psi _{jm}&=&\sqrt {\left( {j\mp m} \right)
\left( {j\pm m+1} \right)}\psi _{jm \pm 1},\\
J_3\psi _{jm}&=&m\psi _{jm},\\
\nonumber K_3\psi _{jm}&=&a_{jm}\psi _{j-1,m}+b_{jm}
\psi _{jm}+a_{j+1,m}\psi _{j+1,m},
\end{eqnarray}
where
 \[a_{jm}\equiv \left( {{{\left( {j^2-m^2} \right)\left( {j^2-\ell _0^2} \right)\left( {\ell _1^2-j^2} \right)} \over {\left(
{2j-1} \right)j^2\left( {2j+1} \right)}}} 
\right)^{{1 \mathord{\left/ {\vphantom {1 2}} \right. \kern-\nulldelimiterspace}
2}},\; \; b_{jm}={{m\ell _0\ell _1} \over {j\left( {j+1} \right)}},\]

\noindent with $\ell _0=j_1-j_2\; \; ,\; \ell _1=j_1+j_2+1$ the labels of the representations.

The representation for $K_1, K_2$ are obtained with the help of the commutation relations.

The Casimir operators are
\begin{equation}
\left( {\bar J^2+\bar K^2} \right)\psi _{jm}=\left( {\ell _0^2+\ell _1^2-1} \right)\psi _{jm},
\end{equation}
\begin{equation}
\bar J\cdot \bar K\psi _{jm}=\ell _0\ell _1\psi _{jm}.
\end{equation}
The representations in the bases $\psi _{jm}$ are irreducible in the following cases
\begin{eqnarray*}
\ell _0&=&j_1-j_2=0,\pm {1 \over 2},\pm 1,\pm {3 \over 2},\pm 2,\ldots,\\
\ell _1&=&j_1+j_2-1=\left| {\ell _0} \right|+1,\left| {\ell _0} \right|+2,\ldots,\\
j&=&\left| {j_1-j_2} \right|,\ldots, j_1+j_2.
\end{eqnarray*}

If we exponentiate the infinitesimal generators we obtain the finite representations of $SO(4,R)$ given in terms of the rotation
angles. An element $U$ of $SO(4,R)$ is given as [13]
\begin{equation}
U\left( {\varphi ,\theta ,\tau ,\alpha ,\beta ,\gamma } \right)=R_3\left( \varphi  \right)R_2\left( \theta  \right)S_3\left(
\tau  \right)R_3\left( \alpha  \right)R_2\left( \beta  \right)R_3
\left( \gamma  \right),
\end{equation}
where $R_2$ is the rotation matrix in the $(x_1x_3)$ plane, $R_3$ the rotation matrix in the $(x_1x_2)$ plane and $S_3$ the
rotation (``boost'') in the $(x_3x_4)$ plane, and
\[0\le \beta ,\psi ,\theta \le \pi \;\;,\;\;0\le \alpha ,\varphi ,\gamma \le 2\pi.
 \]

In the basis $\psi _{jm}$ the action of $S_3$ is as follows:
\begin{equation}
S_3\left( \tau  \right)\psi _{jm}=
\sum\limits_{j'} {d_{j'jm}^{j_1j_2}}\left( \tau  \right)\psi _{j'm},
\end{equation}
where
\begin{equation}
d_{j'jm}^{j_1j_2}\left( \tau  \right)=\sum\limits_{m_1m_2} {\left\langle {{j_1j_2m_1m_2}} \mathrel{\left | {\vphantom
{{j_1j_2m_1m_2} {jm}}} \right. \kern-\nulldelimiterspace} {{jm}} \right\rangle }e^{-i\left( {m_1-m_2} \right)\tau }\left\langle
{{j_1j_2m_1m_2}} \mathrel{\left | {\vphantom {{j_1j_2m_1m_2} {j'm}}} \right. \kern-\nulldelimiterspace} {{j'm}} \right\rangle
\end{equation}
is the Biedenharn-Dolginov function [14].

From this function the general irreducible representations of the operator $U$ in terms of rotation angles is [13]:
\begin{equation}
U\left( {\varphi ,\theta ,\tau ,\alpha ,\beta ,\gamma } \right)\psi _{jm}=\sum\limits_{j'm'} {D_{j'm'jm}^{j_1j_2}}\left( {\varphi
,\theta ,\tau ,\alpha ,\beta ,\gamma } \right)\psi _{j'm'},
\end{equation}
where
\begin{equation}
D_{j'm'jm}^{j_1j_2}\left( {\varphi ,\theta ,\tau ,\alpha ,\beta ,\gamma } 
\right)=\sum\limits_{m''} {D_{m'm''}^{j'}\left(
{\varphi ,\theta ,0 } \right)}d_{j'jm''}^{j_1j_2}
\left( \tau  \right)D_{m''m}^{j}\left( {\alpha ,\beta ,\gamma } \right).
\end{equation}

We now give some particular values of these representations. In the case of spin $j={1 \mathord{\left/ {\vphantom {1 2}} \right.
\kern-\nulldelimiterspace} 2}$ we know
\[ R_3\left( \alpha  \right)R_2\left( \beta  \right)R_3\left( \gamma  \right)=\left( {\matrix{{\cos {\beta  \over 2}e^{i{{\alpha
+\gamma } \over 2}}}&{i\sin {\beta  \over 2}e^{-i\left( {{{\gamma -\alpha } \over 2}} \right)}}\cr {i\sin{\beta 
\over 2}e^{i{{\gamma -\alpha } \over 2}}}&{\cos {\beta  \over 2}e^{-i\left( {{{\alpha +\gamma } \over 2}} \right)}}\cr }} \right)
\]

Introducing the new parameters $\alpha +\gamma =\delta \;\;,\;\;\gamma -\alpha =\eta $ and the variables
\begin{eqnarray*}
x_1=\cos {\beta  \over 2}\cos {\delta  \over 2}\;\;,\;\;x_2=\cos {\beta  \over 2}\sin{\delta  \over 2},\\
x_3=\sin {\beta  \over 2}\sin {\eta  \over 2}\;\;,\;\;
x_4=\sin {\beta  \over 2}\cos {\eta  \over 2},
\end{eqnarray*}
we have
\begin{equation}
R_3\left( \alpha  \right)R_2\left( \beta  \right)R_3\left( \gamma  \right)=\left( {\matrix{{x_1+ix_2}&{x_3+ix_4}\cr
{-x_3+ix_4}&{x_1-ix_2}\cr
}} \right).
\end{equation}
Similarly we have
\begin{equation}
R_3\left( \varphi  \right)R_2\left( \theta  \right)S_3\left( \tau  \right)=\left( {\matrix{{y_1+iy_2}&{y_3+iy_4}\cr
{-y_3+iy_4}&{y_1-iy_2}\cr
}} \right),
\end{equation}
with
\begin{eqnarray*}
y_1=\cos {\theta  \over 2}\cos {{\varphi +\tau } \over 2}\;\;,\;\;y_2=\cos {\theta  \over 2}\sin {{\varphi +\tau } \over 2},\\
y_3=\sin {\theta  \over 2}\sin {{\tau -\varphi } \over 2}\;\;,\;\;y_4=\sin {\theta  \over 2}\cos {{\tau -\varphi } \over 2}.
\end{eqnarray*}

For the Biedenharn-Dolginov function we have some particular values [15]
\begin{eqnarray*}
&&d_{jmm}^{\left[ {j_+,0} \right]}\left( \tau  \right)=i^{j-m}2^j\sqrt {2j+1}\Gamma \left( {j+1} \right)\times \\
&&\times \left( {{{\Gamma \left(
{m+{3 \over 2}} \right)\Gamma \left( {j_+-m+1} \right)\Gamma \left( {j_+-j+1} \right)\Gamma \left( {j+m+1} \right)} \over {\Gamma
\left( {{3 \over 2}} \right)\Gamma \left( {j_++m+2} \right)\Gamma \left( {j_++j+2} \right)\Gamma \left( {j-m+1} \right)\Gamma
\left( {m+1} \right)}}} \right)^{{1 \over 2}}
\end{eqnarray*}

\begin{equation}
\times \left( {\sin \tau } \right)^{j-m}C_{j_+-j}^{j+1}\left( {\cos \tau } \right),
\end{equation}
where $j_+\equiv j_1+j_2\;\;,\;\;j_-=j_1-j_2=0$, and $C_n^\nu \left( {\cos \tau } \right)$ are the Gegenbauer (ultraspherical)
polynomials which are related to the Jacobi polynomials by 
\[C_n^\nu \left( {\cos \tau } \right)={{\Gamma \left( {\nu +{3 \over 2}} \right)\Gamma \left( {2\nu +n} \right)} \over {\Gamma
\left( {2\nu } \right)\Gamma \left( {\nu +n+{1 \over 2}} \right)}}P_n^{\left( {\nu -{1 \over 2},\nu -{1 \over 2}} \right)}\left(
{\cos \tau } \right),
\]
from which it can be deduced that
\begin{equation}
d_{000}^{\left[ {j_+,0} \right]}\left( \tau  \right)={1 \over {j_++1}}{{\sin \left( {\left( {j_++1} \right)\tau } \right)} \over
{\sin \tau }}.
\end{equation}

From the asymptotic relations of $C_n^\nu \left( {\cos \tau } \right)$ it can be proved
\begin{equation}
d_{jmm}^{\left[ {j_+,0} \right]}\left( \tau  \right)\mathrel{\mathop{\kern0pt\longrightarrow}\limits_{j_+\to \infty }}{{i^{j-m}}
\over {j_+^{m+1}}}{{\cos \left[ {\left( {j_++1} \right)\tau -{1 \over 2}\left( {j+1} \right)\pi } \right]} \over {\left( {\sin
\tau } \right)^{m+1}}}.
\end{equation}

\section{Relativistic spin network in 4-dimensions}

We address ourselves to the Barrett-Crane model that generalized Penrose's spin networks from three dimensions to four dimensions
[16]. They characterize the geometrical properties of 4-simplices, out of which the tesselation of the 4-dimensional manifold is made,
and then attach to them the representations of $SO(4)$.

A geometric 4-simplex in Euclidean space is given by the embedding of an ordered set of 5 points in $R^4(0,x,y,z,t)$ which is
required to be non-degenerate (the points should not lie in any hyperplane). Each triangle in it determines a bivector
constructed out of the vectors for the edges. Barrett and Crane proved that classically, a geometric 4-simplex in Euclidean space
is completely characterized (up to parallel translation an inversion through the origin) by a set of 10 bivectors $b_i$, each
corresponding to a triangle in the 4-simplex and satisfying the following properties:
 \begin{enumerate}
\item [i)]  
the bivector changes sign if the orientation of the triangle is changed;

\item [ii)] each bivector is simple, i.e. is given by the wedge product of two vectors for the edges;

\item [iii)]  if two triangles share a common edge, the sum of the two bivector is simple;

\item [iv)]  the sum (considering orientation) of the 4 bivectors corresponding to the faces of a tetrahedron is zero;

\item [v)]  for six triangles sharing the same vertex, the six corresponding bivectors are linearly independent;

\item [vi)]  the bivectors (thought of as operators) corresponding to triangles meeting at a vertex of a tetrahedron satisfy tr $b_1\left[
{b_2,b_3}
\right]>0$ i.e. the tetrahedron has non-zero volume.
 \end{enumerate}
Then Barrett and Crane define the quantum 4-simplex with the help of bivectors thought as elements of the Lie algebra $SO(4)$,
associating a representation to each triangle and a tensor to each tetrahedron. The representations chosen should satisfy the
following conditions corresponding to the geometrical ones:
\begin{enumerate}
\item [i)] different orientations of a triangle correspond to dual representations;

\item [ii)] the representations of the triangles are ``simple'' representations of $SO(4)$, i.e. $j_1=j_2$;

\item [iii)] given two triangles, if we decompose the pair of representations into its Clebsch-Gordan series,
the tensor for the tetrahedron is decomposed into summands which are non-zero only for simple representations;

\item [iv)] the tensor for the tetrahedron is invariant under $SO(4)$. 

Now it is easy to construct an amplitude for the quantum
4-simplex. The graph for a relativistic spin network is the 1-complex, dual to the boundary of the 4-simplex, having five
4-valent vertices (corresponding to the five tetrahedra), with each of the ten edges connecting two different vertices
(corresponding to the ten triangles of the 4-simplex each shared by two tetrahedra). Now we associate to each triangle (the
dual of which is an edge) a simple representation of the algebra $SO(4)$ and to each tetrahedra (the dual of which is a
vertex) a intertwiner; and to a 4-simplex the product of the five intertwiner with the indices suitable contracted, and the
sum for all possible representations. The proposed state sum suitable for quantum gravity for a given triangulation
(decomposed into 4-simplices) is 
 \vspace{-0,2cm}
\end{enumerate}
\begin{equation}
Z_{BC}=\sum\limits_J {\prod\limits_{{\rm triang.}} {A_{{\rm tr}}}}\prod\limits_{{\rm tetrahedra}} {A_{{\rm
tetr.}}}\prod\limits_{4-{\rm simplices}} {A_{{\rm simp.}}}
\end{equation}
where the sum extends to all possible values of the representations $J$.

\section{The triple product in $R^4$}

Before we apply the representation theory developed in previous sections to the Barrett-Crane model we introduce some geometrical
properties based in the triple product that generalizes the vector (cross) product in $R^3$. Given three vectors in $R^4$, we
define the triple product:
\begin{eqnarray}
\nonumber u\wedge v\wedge w&=&-v\wedge u\wedge w=-u\wedge w\wedge v=-w\wedge v\wedge u=v\wedge w\wedge u=\\
\nonumber &=&w\wedge u\wedge v,\\
u\wedge u\wedge v&=&u\wedge v\wedge u=v\wedge u\wedge u=0.
\end{eqnarray}

If the vectors in $R^4$ have cartesian coordinates \[u=\left( {u_1,u_2,u_3,u_4} \right)\;,\;v=\left( {v_1,v_2,v_3,v_4}
\right)\;,\;w=\left( {w_1,w_2,w_3,w_4} \right),\] we define an orthonormal basis in $R^4$
\[ \hat \imath=\left( {1,0,0,0} \right)\;,\;\;\hat \jmath=\left( {0,1,0,0} \right)\;,\;\;\hat k=\left( {0,0,1,0}
\right)\;,\;\;\hat
\ell =\left( {0,0,0,1} \right).
\]
The triple product of these vectors satisfies

\[\hat \imath\wedge \hat \jmath\wedge \hat k=-\hat \ell \;,\;\;\hat \jmath\wedge \hat k\wedge \hat \ell =\hat \imath\;,\;\;\hat
k\wedge \hat \ell \wedge \hat \imath=-\hat \jmath\;,\;\;\hat \imath\wedge \hat \jmath\wedge \hat \ell =\hat k\;.\]

In coordinates the triple product is given by the determinant
\begin{equation}
u\wedge v\wedge w=\left| {\matrix{{\hat \imath}&{\hat \jmath}&{\hat k}&{\hat \ell }\cr
{u_1}&{u_2}&{u_3}&{u_4}\cr
{v_1}&{v_2}&{v_3}&{v_4}\cr
{w_1}&{w_2}&{w_3}&{w_4}\cr
}} \right|.
\end{equation}

The scalar quadruple product is defined by
\begin{eqnarray}
\nonumber a\cdot \left( {b\wedge c\wedge d} \right)&=&\left| {\matrix{{a_1}&{a_2}&{a_3}&{a_4}\cr
{b_1}&{b_2}&{b_3}&{b_4}\cr
{c_1}&{c_2}&{c_3}&{c_4}\cr
{d_1}&{d_2}&{d_3}&{d_4}\cr
}} \right|=\left[ {abcd} \right]=-\left[ {abdc} \right]=\\ 
&=&-\left[ {acbd} \right]=\left[ {acdb} \right] \mbox {and so on}.
\end{eqnarray}
It follows: $a\cdot a\wedge b\wedge c=b\cdot a\wedge b\wedge c=c\cdot a\wedge b\wedge c=0$.

We can use the properties of the three vector for the description of the 4-simplex. Let $\{ 0,x,y,z,t\}$ be the 4-simplex in
$R^4$. Two tetrahedra have a common face

$\left\{ {0,x,y,z} \right\}\cap \left\{ {0,x,y,t} \right\}=\left\{ {0,x,y} \right\}.$

Each tetrahedron is embedded in an hyperplane characterized by a vector perpendicular to all the vectors forming the tetrahedron.
For instance,
 
$\{ 0,x,y,z\}$ is characterized by $a=x\wedge y\wedge z,$

$\{ 0,x,y,t\}$ is characterized by $b=x\wedge y\wedge t.$

\begin{center} 
\includegraphics{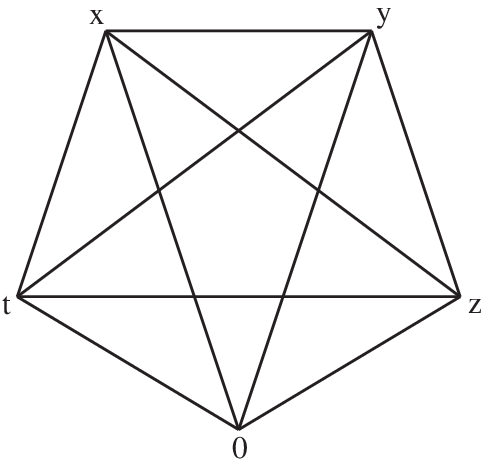}
\end{center}

The vector $a$ satisfies $a\cdot x=a\cdot y=a\cdot z=0,$

the vector $b$ satisfies $b\cdot x=b\cdot y=b\cdot t=0.$

The triangle $\{0,x,y\}$ shared by the two tetrahedra is characterized by the bivector $x\wedge y$. The plane where the triangle
is embedded is defined by the two vectors $a,b$, forming the angle $\phi $, given by 
\[ \cos \phi =a\cdot b. \]
The bivector $a\wedge b$ can be calculated with the help of trivectors as 
\[ a\wedge b=\left[ {x\,y\,z\,t} \right]^*\left( {x\wedge y} \right).\]
Obviously $a\wedge b$ is perpendicular to $x\wedge y$ 
\begin{equation}
\left\langle {a\wedge b,x\wedge y} \right\rangle =\left( {a\cdot x} \right)\left( {b\cdot y} \right)-\left( {a\cdot y}
\right)\left( {b\cdot x} \right)=0.
\end{equation}
For completness we add some useful properties of bivectors in $R^4$. The six components of a bivector can be written as 
$$\begin{array}{lll} 
B_{\mu\nu }=x_\mu y_\nu -x_\nu y_\mu \;\quad ,\quad \;&\mu ,\nu =1,2,3,4\;\quad ,\quad \; &B=\left( {\bar J,\bar K} \right),\\
J_1=\left( {x_2y_3-x_3y_2} \right)\;\quad ,\quad \;&J_2=\left( {x_3y_1-x_1y_3}
\right)\;\quad ,\quad \;&J_3=\left( {x_1y_2-x_2y_1} \right),\\
K_1=\left( {x_1y_4-x_4y_1} \right)\quad ,\quad &K_2=\left( {x_2y_4-x_3y_1} \right)\quad ,\quad &K_3=\left(
{x_3y_4-x_4y_1} \right).
\end{array}$$
The six components of the dual of a bivector are 

\noindent $^*B=\left( {\bar K,\bar J} \right)\;,\quad ^*B_{\alpha \beta }={1 \over 2}b_{\mu \nu }\,\varepsilon _{\mu \nu \alpha
\beta }.$

We take the linear combinations of $\bar J,\bar K$
\begin{equation}
\bar M={1 \over 2}\left( {\bar J+\bar K} 
\right),\quad \quad \bar N={1 \over 2}\left( {\bar J-\bar K} \right).
\end{equation}
 They form the bivector $\left( \bar M,\bar N \right)$, whose dual is:
\begin{equation}
^*\left( {M,N} \right)=\left( {M,-N} \right),
\end{equation}
therefore $\bar M$ can be considered the self-dual part, $\bar N$ the antiselfdual part of the bivector $(\bar M,\bar N)$.
$\bar M$ and $\bar N$ coincides with the basis for the irreducible tensor representations of section 3. The norm of the
bivectors can be explicitly calculated.
\begin{eqnarray}
\nonumber\left\| B \right\|^2&=&\left\langle {B,B} \right\rangle =J^2+K^2=\left\| x \right\|^2\left\| y \right\|^2-\left| {x,y}
\right|^2=\\
&=&\left\| x \right\|^2\left\| y \right\|^2\sin ^2\phi 
(x,y)=4Area^2\left\{ {0,x,y} \right\},
\end{eqnarray}
\begin{equation}
\left\| {^*B} \right\|^2=\left\langle {^*B,^*B} \right\rangle =J^2+K^2=\left\| B \right\|^2.
\end{equation}
Finally, the scalar product of two vectors in $R^4$ can be expressed in terms of the corresponding $U\left( {2,C\!\!\!\!/}
\right)$ matrices

\medskip

Let $X\Leftrightarrow \left( {\matrix{{x_1+ix_2}&{x_3+ix_4}\cr
{-x_3+ix_4}&{x_1-ix_2}\cr
}} \right),\quad \quad Y\Leftrightarrow 
\left( {\matrix{{y_1+iy_2}&{y_3+iy_4}\cr
{-y_3+iy_4}&{y_1-iy_2}\cr
}} \right).$

Then  
\begin{equation}
Tr\left( {X^+Y} \right)=x_1y_1+x_2y_2+x_3y_3+x_4y_4.
\end{equation}

\section{Evaluation of the spin sum for the relativistic spin network}

The five tetrahedra in the 4--simplex are numbered by $k=1,2,3,4,5$ and the triangles are indexed by the pair $k,l$ of tetrahedra
which intersect on the triangle $kl$. To each triangle we associate a simple 
representation of $SO(4)$ labelled by $j_{kl}$, that
corresponds to the same spin for each part of the $SU(2)\otimes SU(2)$ group. 
The matrix representing an element $g\in SU(2)$ in
the irreducible representation of spin $j_{kl}$ belonging to a triangle is denoted by $\rho _{kl}(g)$. An element $h_k\in
SU(2)$ is assigned to each tetrahedron $k$. The invariant $I$ is defined by integrating a function of these variables over each
copy of $SU(2)$:
\begin{equation}
I=(-1)^{\sum\limits_{k<\ell } {2j_{k\ell }}}\int\limits_{h\in SU(2)^5} {\prod {{\rm Tr}\rho _{k\ell }}}\left( {h_kh_\ell ^{-1}}
\right)
\end{equation}
The geometrical interpretation of this formula given by Barrett [17] is that since the manifold $SU(2)$ is isomorphic to $S^3$,
each variable $h\in SU(2)$ can be regarded as a unit vector in $R^4$. This unit vector can be identified with the vector
perpendicular to the hyperplane where the tetrahedron is embedded. The two variables $h_k, h_l$ correspond in this picture to the
two vectors $a,b$ that were defined in the last section.

In our opinion there is some disagreement between the conditions given in Ref. [16] and the application of formula (2.1) in
Ref.[17]. In the former paper an irreducible representation of $SO(4)$ with two labels $j_1=j_2$ is assigned to each triangle in
the 4--simplex. In the last paper, a representation of $SU(2)$ is assigned to each triangle. Therefore we have the standard values
for the trace of a general representation of the group $SU(2)$ with spin $j$, namely, $\sin \left( {2j+1} \right)\phi /\sin \phi
$ (Formula 4.1 of Ref. 17).

The disagreement can be avoided if one takes the trace with respect to the irreducible representation of $SO(4)$ as described in
Sections 3 and 4, where the parameters of the group $SO(4)$ are the $3+3$ cartesian independent coordinates of the two unit
vectors $h_k, h_{\ell}$, as defined before, or the 6 rotation angles of formula (28). In the last case we choose a system of
reference for $R_4$ such that one unit vector corresponding to $h_k$, say $a$, has components $(0,0,0,1)$ and the other one $h_{\ell}$, say
$b$, is located in the plane $(x_3 x_4)$ forming an angle $\tau$ with the first vector. In this particular situation all the
rotation angles $\alpha =\beta =\gamma =\vartheta =\varphi =0$ and the representation is restricted to $S_3(\tau)$.

From (31) and (32) the general element representation of $SO(4)$ is restricted to
\begin{equation}
D_{j'm'jm}^{j_1j_2}(0,0,\tau ,0,0,0)=d_{j'jm}^{j_1j_2}(\tau )\equiv d_{j'jm}^{\left[ {j_+,j_-} \right]}(\tau ).
\end{equation}

In the case of a simple representations of $SO(4)$ $j_-=j_1-j_2=0$, and the trace becomes
\begin{equation}
{\rm tr}D_{j'm'jm}^{\left[ {j_+,0} \right]}(\tau) =\sum\limits_{j=0}^{j_+} {\sum\limits_{m=-j}^j {d_{jjm}^{\left[ {j_+,0} \right]}(\tau
)}}.
\end{equation}

Obviously this expression does not coincide with formula (4.1) of Ref. [17] except in the term
\[d_{000}^{\left[ {j_+,0} \right]}(\tau )={1 \over {j_++1}}{{\sin \left( {( {j_++1} )\tau } \right)} \over
{\sin \tau }}
\mbox {, Ref [15], (IV.2.9).} \]

For other values of the Biedenharn-Dolginov function we can use the asymptotic expression (37) for $m=j$. With this formula it is
still possible to give an geometrical interpretation of the probability amplitude encompassed in the trace. In fact, the spin
dependent factor appearing in the exponential of (37)
\begin{equation}
e^{i\left( {2j_{k\ell }+1} \right)\tau _{k\ell }},
\end{equation}
corresponding to two tetrahedra $k \ell$ intersecting the triangle $k\ell$, can be interpreted as the product of the angle
between the two vectors $h_k, h_{\ell}$ perpendicular to the triangle and the area $A_{k\ell}$ of the intersecting triangle.

For the proof we identify the component of the antisymmetrie tensor $\left( {\bar J,\bar K} \right)$ with the components of the
infinitesimal generators of the $SO(4)$ group
\[J_{\mu \nu }\equiv i\left( {x_\mu {\partial  \over {\partial x_\nu }}-x_\nu {\partial  \over {\partial x_\mu }}} \right).
\]
From (43) and (45) we have
$\left\| {B} \right\|^2=4\left( {A_{k\ell }} \right)^2=2\left( {\bar M^2+\bar N^2} \right)$

But $\bar M^2$ and $\bar N^2$ are the Casimir operators of the $SU(2) \otimes SU(2)$ group with eigenvalues $j_1\left( {j_1+1}
\right)$ and $j_2\left( {j_2+1}\right)$.

For large values of $j_1=j_2=j_{k \ell}$ we have
\[2\left( {\bar M^2+\bar N^2} \right)
\cong 4j_{k\ell }^2+4j_{k\ell }+1=\left( {2j_{k\ell }+1} \right)^2,\]
therefore ${1 \over 2}\left( {2j_{k\ell }+1} \right)=A_{k\ell }$ where $A_{k\ell }$ is the area of the triangle characterized by
the two vectors $h_k$ and $h_{\ell}$ and $j_{k \ell}$ is the spin corresponding to the representation $\rho _{k\ell }$ associated
to the triangle $kl$. Substituting this result in (51) we obtain the assimptotic value of the amplitude given by Barrett and
Williams [18]. 

\acknowledgements
The authors would like to express their gratitude to Professor Bruno Gruber for inviting them to the Colloquium ``Symmetries in Science
XIII''. One of the authors (M.~L.) wants to extend his gratitude to the Director of the Institut f\" ur theoretische Physik of the
University of T\" ubingen, where part of this work was done, and also to the Ministerio de Ciencia y Tecnolog\'{\i}a (Spain) for the finantial
support under grant BFM-2000--0357

\end{article}
\end{document}